\AtBeginDocument{%
  }
    
\documentclass[sigconf, 9pt]{acmart}

\usepackage{amsmath,amssymb,amsfonts}
\usepackage{algpseudocode}
\usepackage{algorithm}
\usepackage{graphicx}
\usepackage{textcomp}
\usepackage{xcolor}
\usepackage{hyperref}
\usepackage{soul}
\usepackage{physics,array}
\usepackage{multirow}
\usepackage{tikz}
\usepackage{adjustbox}
\usepackage{geometry}
\usepackage{changepage}
\usepackage{pifont}
\usepackage{amsmath}
\usepackage{amssymb}
\usepackage{caption}

\newcommand*\whitecircled[1]{\tikz[baseline=(char.base)]{
            \node[shape=circle,draw,fill=white,inner sep=0.5pt] (char) {\textcolor{black}{#1}};}}

\setcopyright{acmlicensed}
\copyrightyear{2018}
\acmYear{2018}
\acmDOI{XXXXXXX.XXXXXXX}

\renewcommand\footnotetextcopyrightpermission[1]{}

\acmBooktitle{Woodstock '18: ACM Symposium on Neural Gaze Detection,
 June 03--05, 2018, Woodstock, NY}
\acmISBN{978-1-4503-XXXX-X/18/06}

\begin{document}
\fancyhead{}


\title{The Q-Spellbook: Crafting Surface Code Layouts and Magic State Protocols for Large-Scale Quantum Computing}
    

\author{Avimita Chatterjee}
\orcid{1234-5678-9012}
\affiliation{%
  \institution{Pennsylvania State University}
  \city{State College}
  \state{PA}
  \country{USA}
}
\email{amc8313@psu.edu}

\author{Archisman Ghosh}
\affiliation{%
  \institution{Pennsylvania State University}
  \city{State College}
  \state{PA}
  \country{USA}
  }
\email{apg6127@psu.edu}

\author{Swaroop Ghosh}
\affiliation{%
  \institution{Pennsylvania State University}
  \city{State College}
  \state{PA}
  \country{USA}
}
\email{szg212@psu.edu}


\begin{abstract}
Quantum error correction is a cornerstone of reliable quantum computing, with surface codes emerging as one of the most prominent methods for protecting quantum information. Surface codes are highly efficient for Clifford gates but require magic state distillation protocols to process non-Clifford gates, such as T gates, which are essential for universal quantum computation. In large-scale quantum architectures capable of correcting arbitrary quantum circuits, it becomes necessary to design specialized surface codes for data qubits and distinct surface codes tailored for magic state distillation. Consequently, such architectures can be organized into data blocks and distillation blocks. 
The system works by having distillation blocks produce magic states and data blocks consume them, causing stalls due to either a shortage or excess of magic states. This bottleneck presents an opportunity to optimize quantum space by balancing data and distillation blocks. While prior research offers insights into selecting distillation protocols and estimating qubit requirements, it lacks a tailored optimization approach.

We present a comprehensive framework for optimizing large-scale quantum architectures, focusing on data block layouts and magic state distillation protocols. 
We evaluate three distinct data block layouts (\textit{compact}, \textit{intermediate}, and \textit{fast}) and four distillation protocols (\textit{15-to-1}, \textit{116-to-12}, \textit{225-to-1}, and \textit{20-to-4}) under key optimization strategies: minimizing tiles, minimizing steps, and achieving a balanced trade-off.
Through a comparative analysis of brute force, dynamic programming, greedy, and random algorithms, we find that brute force delivers true optimal results, while the greedy algorithm deviates by only $7\%$ for minimizing steps and dynamic programming matches brute force in tile minimization. We observe that total steps increase with the number of columns, while total tiles scale with the number of qubits. Finally, we propose a generalized heuristic to help users select well-suited algorithms tailored to their specific objectives, enabling scalable, efficient, and adaptable quantum architectures.
\end{abstract}

\maketitle

\section{Introduction} \label{sec:introduction}

Quantum error correction (QEC)~\cite{preskill1998reliable, terhal2015quantum, campbell2017roads, devitt2013quantum, lidar2013quantum, roffe2019quantum} is a cornerstone of modern quantum computing, addressing quantum systems' inherent noise and decoherence challenges~\cite{clerk2010introduction} by exponentially suppressing errors. This capability is essential for achieving error rates low enough to enable practical quantum computation. The surface code~\cite{fowler2012surface, kitaev2003fault} stands out as one of the most extensively studied and implemented methods among the various QEC strategies. It is particularly appealing due to its reliance on local interactions, a high fault-tolerance threshold of approximately 0.7\%~\cite{google2023suppressing, acharya2024quantum}, and its ability to support universal quantum computation when combined with a magic state factory~\cite{bravyi2012magic, litinski2019magic}.

A well-established low-overhead approach to fault-tolerant quantum computation leverages the Clifford + T gate formalism~\cite{kliuchnikov2012fast, brown2017poking, litinski2018lattice}. In this model, Clifford gates can be executed fault-tolerantly using surface codes, while T gates, which are non-Clifford gates, require injection through magic state distillation protocols~\cite{haah2018codes, bravyi2005universal, jones2013multilevel, fowler2013surface}. $T$ gates are particularly resource-intensive. While Clifford gates can be executed entirely, the implementation of non-Clifford or $T$ gates requires the consumption of a specialized quantum resource known as a magic state, $|0\rangle + e^{i\pi/4}|1\rangle$ \cite{litinski2018quantum}. However, only faulty (i.e., undistilled) magic states can be initially prepared. To achieve the high fidelity needed for large-scale quantum computation, a complex and resource-intensive procedure called magic state distillation is employed. This paradigm has given rise to Pauli-based computation~\cite{gottesman1998heisenberg}, a framework that optimizes quantum circuits by commuting and pruning Clifford gates while isolating circuit blocks consisting of non-Clifford gates and measurements for targeted error correction~\cite{litinski2019game}.

\subsection{Background}
\begin{figure*}
    \centering
    \includegraphics[width=\linewidth]{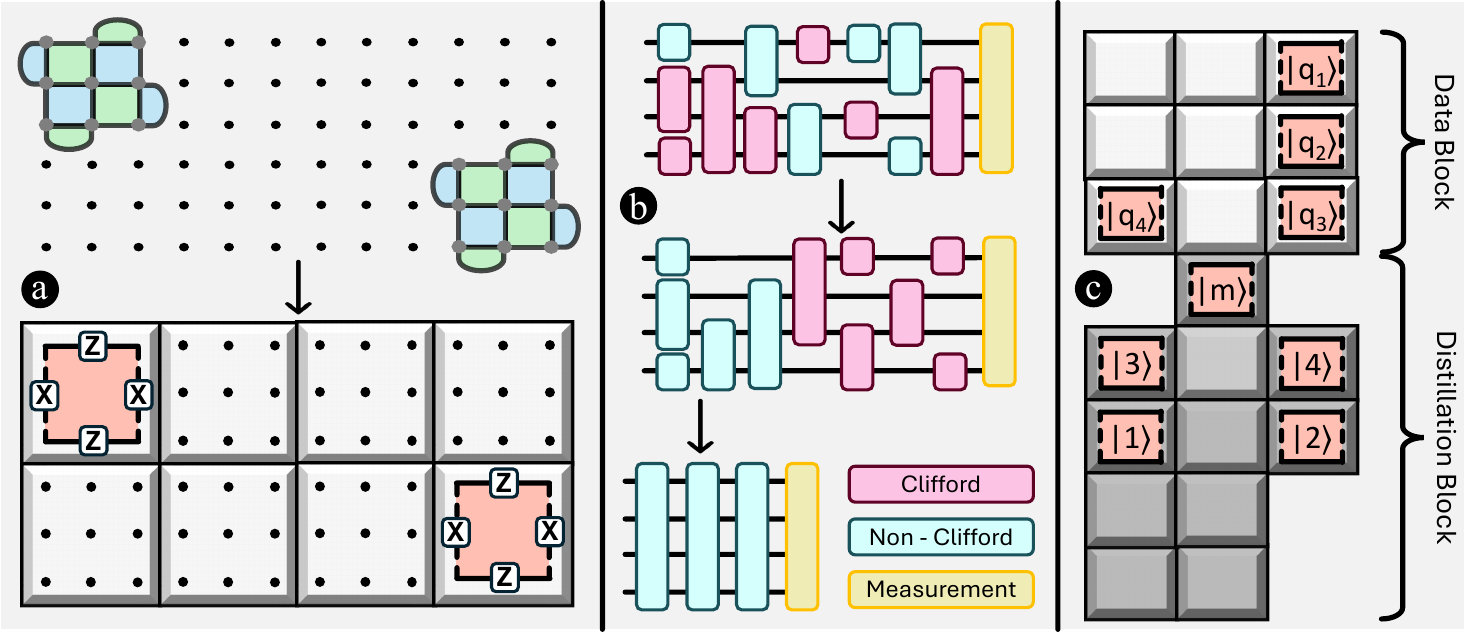}
    \caption{Mapping Quantum Circuits to a Surface Code Framework: 
    (a) The top panel shows a quantum space with physical qubits and two distance-three surface codes, where green and blue surfaces represent Z- and X-stabilizers. The bottom panel divides the space into tiles, with orange patches indicating surface codes and stabilizer boundaries. 
    (b) The first circuit includes Clifford (pink), non-Clifford (cyan), and measurement (yellow) gates, optimized using Pauli-based formalism to eliminate Clifford gates. The final circuit consists of non-Clifford gates and measurements. 
    (c) The optimized circuit is mapped into a quantum space, with light gray tiles forming a `compact' data block and dark gray tiles representing the `15-to-1' distillation block for magic state distillation. Orange patches within each block correspond to surface codes for data qubits and distillation, with empty tiles reserved for ancillary qubits or dynamic computation.
}
    \label{fig:whole_architecture}
\end{figure*}
Fig.~\ref{fig:whole_architecture} illustrates the overall quantum space and the integration of surface codes alongside data qubits and those used in the distillation protocol~\cite{litinski2019magic}. In Fig.~\ref{fig:whole_architecture} \whitecircled{a} (top), the quantum space is depicted with physical qubits and two distance-three surface codes, where the data qubits align with the physical qubits. The green surfaces represent Z-stabilizers, while the blue surfaces correspond to X-stabilizers. Fig.~\ref{fig:whole_architecture} \whitecircled{a} (bottom) presents the conventional representation of this quantum space in the literature. The space is partitioned into tiles, each containing an equal number of physical qubits. Specifically, each gray tile comprises a number of qubits equal to the square of the surface code’s distance. For the distance-three surface code used here, each tile thus consists of nine qubits. The tile corresponding to the surface code placement features an orange patch representing the surface code itself. This patch is delineated by opposing straight and dotted boundaries, where the dotted side indicates the presence of protruding Z-stabilizers and the straight side corresponds to the protruding X-stabilizers of the surface code. For simplicity, subsequent representations will omit the explicit depiction of individual qubits within each tile and will not label the X- and Z-stabilizer sides on the patches.

The first circuit of Fig.~\ref{fig:whole_architecture} \whitecircled{b} illustrates the initial circuit requiring corrections. This circuit consists of four qubits and includes a mix of Clifford gates (depicted in pink), non-Clifford gates (depicted in cyan), and measurement operations (depicted in yellow) applied to all four qubits at the end. Before this circuit can be mapped onto the plane of surface codes for processing, it must undergo a series of transformations. When converting such an arbitrary quantum circuit into the Pauli-based computation formalism, we begin by decomposing the circuit into the Clifford+T gate set, which consists of the gates \(H\), \(S\), \(T\), and \(CX\). These gates are then transformed into sequences of Pauli rotations, expressed as \(e^{i\theta P}\), where \(P \in \{I, X, Y, Z\}\) and \(\theta\) represents the rotation angle. The standard decompositions for these gates are as follows:
$H = Z\left(\frac{\pi}{2}\right) X\left(\frac{\pi}{4}\right) Z\left(\frac{\pi}{2}\right)$, $S = Z\left(\frac{\pi}{4}\right)$, $T = Z\left(\frac{\pi}{8}\right)$, and $CX = ZX\left(\frac{\pi}{4}\right) ZI\left(-\frac{\pi}{4}\right) IX\left(-\frac{\pi}{4}\right)$.


Using these transformations, non-Clifford gates can be optimized by commuting them through Clifford gates. Clifford Pauli product rotations, represented as \(e^{i\theta P}\) with \(\theta = \pm\frac{\pi}{4}\), can be commuted past non-Clifford Pauli product rotations, represented as \(e^{i\theta' P'}\) with \(\theta' = \pm\frac{\pi}{8}\). The commutation rules are as follows:
If \(P'P = P'P\), i.e., the operators commute, then \(P\) can pass \(P'\) without affecting the operator.
If \(P'P = -P'P\), i.e., the operators anti-commute, commuting \(P\) past \(P'\) introduces a phase factor \(i\), transforming \(P'\) into \(iP'\) for the non-Clifford operator. This produces an intermediate circuit similar to that of the second circuit in Fig.~\ref{fig:whole_architecture} \whitecircled{b}. Lastly, measurement operations are optimized by absorbing Clifford gates into the measurement operators. This process effectively removes all Clifford gates from the circuit, resulting in a simplified computation. Consequently, the overall process is streamlined, resulting in a final circuit, similar to the third circuit in Fig.~\ref{fig:whole_architecture} \whitecircled{b}, that consists solely of columns of non-Clifford gates and measurement operations.

Fig.~\ref{fig:whole_architecture} \whitecircled{c} shows the final mapping of the entire circuit onto a quantum space with surface codes for error correction. The light gray tiles represent the data block, which is of the `compact' type and is dedicated to the four data qubits, while the dark gray tiles correspond to the distillation block, used exclusively for magic state distillation. Magic state distillation plays a crucial role in mitigating errors from non-Clifford gates within a surface code framework. Each orange patch on the grid represents a surface code, with patches in the data block corresponding to data qubits and patches in the distillation block dedicated to magic state distillation. The empty tiles without patches are utilized during computation or to accommodate ancillary qubits. Therefore, the total number of qubits required includes all qubits in the tiles, not just those with patches. 
To correct a circuit, it must be placed on a plane with a defined data block and distillation block. The distillation block generates magic states, while the data block consumes them, creating a dynamic interplay. A distillation protocol is an \( N \)-to-\( K \) protocol, where \( N \) is the number of input magic states and \( K \) is the number of distilled states produced. Multiple distillation blocks may use the same or different protocols, but only one data block is allowed. In this figure, a `compact' data block is used with a `15-to-1' distillation protocol.

The entire system requires a specific number of time steps to process the circuit. Each tile corresponds to a total of $2d^2 - 1$ qubits in a distance-$d$ surface code, out of which $d^2$ qubits are for data qubits, and each time step approximately corresponds to $d$ code cycles, with each code cycle lasting $1~\mu\text{s}$. In this paper, all calculations and results will be presented in terms of the number of tiles and steps.

\subsection{Motivation}
\begin{figure}
    \centering
    \includegraphics[width=\linewidth]{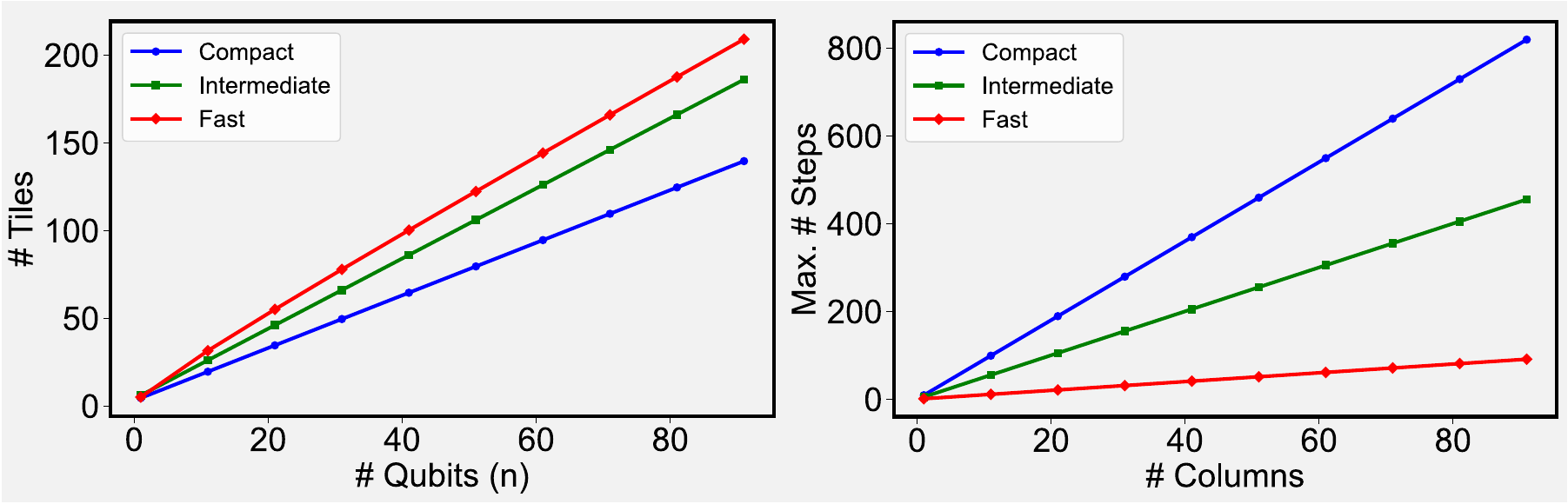}
    \caption{Trade-off Between Space and Time in Data Block Types: 
    The left figure shows how the number of tiles increases with qubits for each data block type (Compact, Intermediate, Fast). The right figure illustrates the increase in maximum steps as the number of columns grows, highlighting the space-time trade-off: compact blocks use less space but take longer to consume a magic state, while fast blocks consume magic states quickly but require more space.
}
    \label{fig:data_block_characteristics}
\end{figure}

\begin{table}[]
\centering
\fontsize{8.5pt}{9.5pt}\selectfont
\caption{Comparison of data blocks, their total tile, and maximum steps usage to consume magic states per circuit column \cite{litinski2019game}.}
\begin{tabular}{c||cc}
\textbf{Data Block} & \# \textbf{Tiles} & \# \textbf{Max. Steps to Consume} \\ \hline \hline
\textbf{Compact}         & $\lfloor 1.5n + 3 \rfloor$       & 9                                   \\
\textbf{Intermediate}    & $\lfloor 2n + 4 \rfloor$       & 5                                   \\
\textbf{Fast}            & $\lfloor 2n + \sqrt{8n + 1} \rfloor$       & 1                                   \\ \hline \hline
\end{tabular}
\label{tab:data_block_details}
\end{table}
\begin{table}[]
\centering
\fontsize{8.5pt}{9.5pt}\selectfont
\caption{Comparison of distillation protocols, including tile usage, steps for magic state production, success probability, and average steps per successful state \cite{litinski2019game}.}
\begin{tabular}{c||cccc}
\textbf{Protocol}  & \# \textbf{Tiles} & \# \textbf{Steps} & \textbf{Succ.\%} & \textbf{Avg. Steps / Succ.} \\ \hline \hline
\textbf{15-to-1}   & 11             & 11             & 99.85               & 11.02                                            \\
\textbf{20-to-4}   & 14             & 17             & 99.80               & 4.26                                             \\
\textbf{116-to-12} & 44             & 99             & 98.85               & 8.35                                             \\
\textbf{225-to-1}  & 176            & 15             & 97.78               & 15.34                                            \\ \hline \hline
\end{tabular}
\label{tab:distillation_block_details}
\end{table}
Table~\ref{tab:data_block_details} shows the data block types, their tile usage, and the maximum steps to consume magic states per circuit column. Fig.~\ref{fig:data_block_characteristics} (left) illustrates how tile count increases with qubits, and (right) shows the increase in steps with columns. This highlights the trade-off: the compact block uses the least space but takes the longest to consume a magic state, while the fast block consumes states quickly but requires more space. 
Table~\ref{tab:distillation_block_details} shows the distillation protocols, including tile usage, steps for magic state production, success probability, and average steps per successful magic state for a system with a physical error rate of \( p = 10^{-4} \). For an \( N \)-to-\( K \) protocol, the success probability is \( P_s = (1 - p)^N \), and the average time per successful magic state is \( \text{Time} = \frac{S_d}{K \cdot P_s} \), where \( S_d \) is the number of steps per round. This quantifies the efficiency and reliability of distillation protocols~\cite{litinski2019magic}.

Analyzing the data from the previously discussed tables reveals that the steps required to consume magic states differ from those needed for their production in distillation protocols. Since the system must wait for a magic state to be consumed before advancing to the next column, it inevitably experiences stalls due to a shortage of magic states or wastage of resources from excess unused magic states generated.
For example, a circuit with 10 qubits that require 1 magic state starts at time 0 with a compact data block and 20-to-4 distillation block. By time 9, the magic state is ready for consumption, but the system remains idle from times 10 to 16. At time 17, 4 magic states are produced, and at time 18, the magic state is consumed. The process takes 18 steps, uses 32 tiles (18 for the data block and 14 for the distillation block), and includes 7 idle steps.
This bottleneck highlights an opportunity to optimize the quantum space by strategically balancing data blocks and distillation blocks for efficient processing. While prior research offers valuable insights into selecting appropriate distillation protocols for a given physical error rate and estimating qubit requirements for data block arrangements, it does not provide a tailored, optimized approach. Existing literature lacks methods for dynamically determining data block layouts, the number of distillation blocks, and the selection of distillation protocols based on the specific requirements of a quantum circuit and user-defined optimization goals, such as minimizing time, space, or balancing both. This gap motivates the need for a framework that overcomes these limitations, enabling adaptive and user-specific optimizations in large-scale quantum architectures.

\subsection{Contributions}
In this work, we present an optimized framework for large-scale quantum architectures that integrates data block layouts with magic state distillation protocols to address circuit-specific requirements and system constraints. 
We focus on three key optimization strategies: (i) \textit{Minimizing tiles}, which reduces qubit usage for resource-constrained systems, (ii) \textit{Minimizing processing steps}, which accelerates computation for time-sensitive applications, and (iii) a \textit{balanced approach}, which achieves a trade-off between resource usage and processing speed. We implement and compare multiple optimization algorithms—\textit{random}, \textit{brute force}, \textit{dynamic programming}, and \textit{greedy algorithms}—to evaluate their performance across the four optimization strategies and identify the most effective heuristic for optimization. Finally, we propose a generalized heuristic to guide users in selecting the most suitable algorithm for a given optimization objective. For minimizing steps, the greedy algorithm serves as a viable alternative to brute force. For minimizing tiles, dynamic programming—and potentially even the random algorithm—can be effective substitutes for brute force. For balanced results, users aiming to prioritize step reduction can leverage the greedy algorithm with balanced optimization, while those focusing on tile reduction can use dynamic programming with balanced optimization.

\textbf{Paper Structure:}
Section~\ref{sec:circuit_gen} describes the systematic generation of circuit parameters, the structured construction of circuits based on these parameters, and the classification of circuits according to their size and complexity. Section~\ref{sec:methods} details the four optimization algorithms —random search, brute force, dynamic programming, and greedy algorithms —and explains their application across different optimization strategies. Section~\ref{sec:evaluation} provides a comparative analysis of these algorithms and ultimately identifying the most effective heuristic. Finally, Section~\ref{sec:conclusion} summarizes the key findings.
\section{Generating Diverse Circuits} \label{sec:circuit_gen}

\begin{figure}
    \centering
    \includegraphics[width=0.8\linewidth]{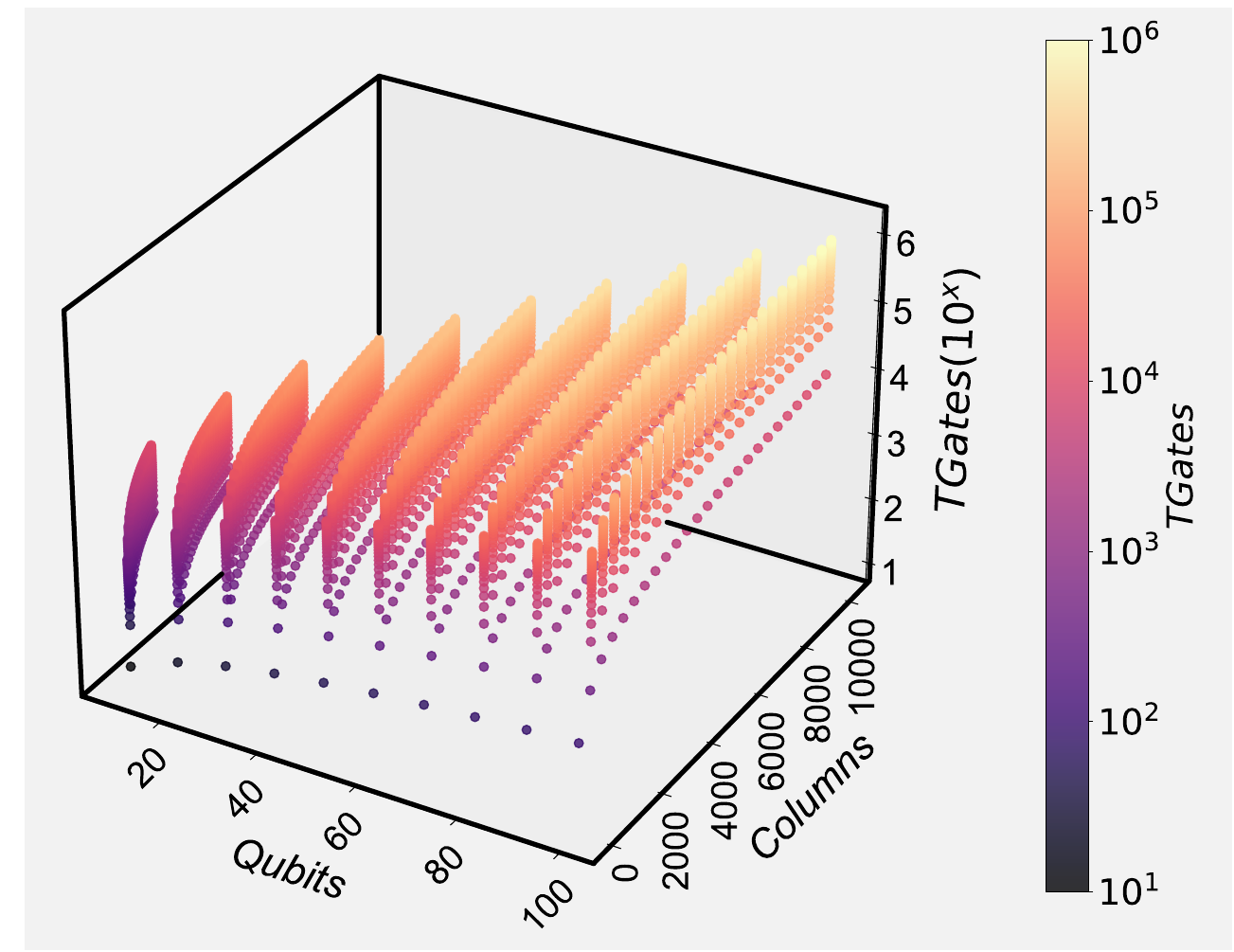}
    \caption{Dataset Distribution Across Circuit Parameters: 
    The dataset spans small circuits to huge ones, encompassing $6,250$ unique quantum circuits. The smallest circuit consists of $10$ qubits, $1$ column, and $10$ T gates, while the largest circuit contains $100$ qubits, $10^4$ columns, and $10^6$ T gates. 
    }
    \label{fig:dataset_representation}
\end{figure}

\textbf{Generation of Circuit Parameters:}
To generate the dataset for this study, we vary three key parameters of each quantum circuit: the number of qubits, the number of columns, and the total number of T gates. We begin by selecting 10 discrete values for the number of qubits, ranging from 10 to 100, using linear spacing. For each chosen number of qubits, the number of columns is varied across 25 discrete values, spanning from a minimum of 1 to a maximum of $100 \times \text{number of qubits}$. 
For each specific combination of qubits and columns, we further generate 25 different cases by varying the total number of T gates from a minimum of $\max(\text{qubits}, \text{columns})$ to a maximum of $\text{qubits} \times \text{columns}$. This dataset spans an extensive range of circuit sizes, covering both small circuits and extremely large circuits. 
Fig.~\ref{fig:dataset_representation} provides a visual representation of each unique circuit.
The chosen range ensures that circuits with varying qubit counts, different depths, and different distributions of T gates are well represented. Additionally, by allowing the number of columns to scale up to $ \times$ the number of qubits, the dataset includes both shallow circuits with high qubit parallelism and deep circuits with fewer qubits but high depth. This diversity allows us to analyze the trade-offs between space (tile usage) and time (computation steps) across different data block and distillation block configurations, ensuring that our study generalizes well to practical large-scale quantum architectures.

\textbf{Structured Circuit Construction:}
After determining the circuit parameters, a random quantum circuit is generated to ensure proper distribution of $T$ gates and variability in gate placement. The process follows these steps:
%
The circuit is initialized as a \( \text{num\_qubits} \times \text{num\_columns} \) grid, with each cell initially containing the identity operation `I', ensuring gates are distributed across qubits (rows) and computational steps (columns).
%
To ensure an even distribution of \( T \) gates, the first step ensures that every qubit (row) and every computational column receive at least one \( T \) gate. This is done by randomly selecting a unique qubit-column pair and placing a randomly chosen Pauli rotation (\( X, Y, Z \)) at that location. 
%
After meeting the minimum requirement, remaining \( T \) gates are randomly placed in available positions (cells with `I') until the total number of \( T \) gates is reached.
%
%
To ensure valid circuits, the number of \( T \) gates must be at least \( \max(\text{num\_qubits}, \text{num\_columns}) \) to ensure each row and column receives at least one gate, and cannot exceed \( \text{num\_qubits} \times \text{num\_columns} \) to prevent over-allocation.
%

\textbf{Circuit Size-Based Classification:}
We use a three-layer classification framework to analyze circuit diversity, categorizing circuits by depth, T-gate density, and qubit system size. Each layer builds on the previous one, with three classifications per layer, resulting in $3 \times 3 \times 3 = 27$ unique categories.

The first layer classifies circuits based on depth, measured by the total number of columns. Using percentile-based thresholds, circuits are categorized as `Shallow ($S$)', `Medium ($M$)', or `Deep ($D$)'. This classification defines the circuit's structural extent.
The second layer evaluates T-gate density, defined as:
$T_{Gate\_Density} = \frac{Total\_T\_Gates}{Qubits \times Columns}$.
Circuits are categorized as `Low ($L$)', `Medium ($M$)', or `High ($H$)' based on their percentile rank. This classification considers the circuit's depth from the first layer, as T-gate density is inversely influenced by both depth and qubits.
The third layer classifies circuits by qubit system size, determined by the total number of qubits. Using the previous classifications, circuits are categorized as `Small ($S$)', `Medium ($M$)', or `Large ($L$)'. This hierarchical process contextualizes the qubit size within the circuit's structure and computational characteristics.

\begin{figure}
    \centering
    \includegraphics[width=1\linewidth]{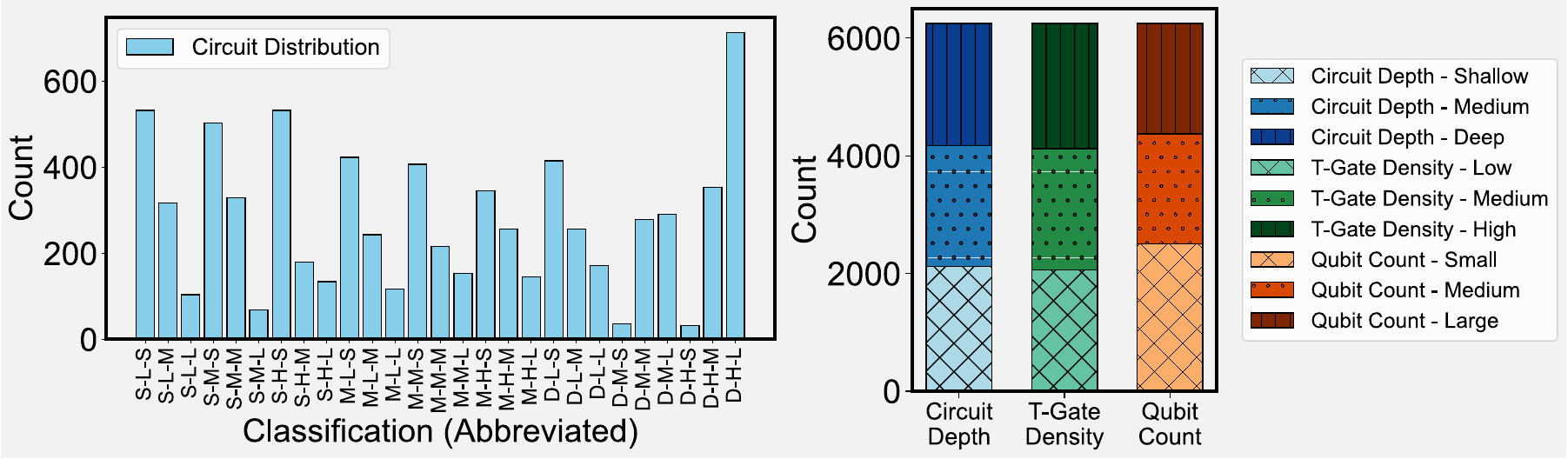}
    \caption{Circuit Classification Distribution: 
    The left figure shows the distribution of circuits across 27 classification categories based on Circuit Depth, T-Gate Density, and Qubit System Size. The right figure displays the distribution across three layers: Circuit Depth, T-Gate Density, and Qubit System Size, with stacked bars representing subcategories in distinct colors and patterns.
}
    \label{fig:circuit_classification}
\end{figure}

To ensure brevity, circuits are labeled using a combination of classifications from the three layers, expressed as $D-T-Q$ or $DTQ$, where $D$ is depth, $T$ is T-gate density, and $Q$ is qubit system size. For example, S-L-S stands for Shallow Depth, Low T-Gate Density, and Small Qubit System, while D-H-L denotes Deep Depth, High T-Gate Density, and Large Qubit System.
Figure~\ref{fig:circuit_classification} (left) shows the distribution of circuits across all 27 classification categories. Fig.~\ref{fig:circuit_classification} (right) illustrates the distribution across the three layers: Circuit Depth, T-Gate Density, and Qubit System Size, with stacked bars indicating the proportion of circuits in each subcategory. The balanced distribution of subcategories within each layer highlights the comprehensive nature of the dataset, ensuring no subcategory disproportionately dominates.

\section{Algorithms and Optimization Strategies} \label{sec:methods}

We focus on three primary optimization objectives: minimizing qubits (tiles), minimizing total execution time (steps), and balancing space and time. Based on the chosen objective, we determine the optimal data block type (compact, intermediate, or fast) and the best combination of distillation protocols (15-to-1, 20-to-4, 116-to-12, and 225-to-1). To achieve these optimizations, we use four approaches: Random, Brute Force, Dynamic Programming, and Greedy algorithms.
%
%
These four approaches were chosen to represent a comprehensive range of optimization strategies, balancing accuracy, efficiency, and applicability. The brute force method guarantees the optimal solution, while dynamic programming offers a more efficient path to similar results. The greedy algorithm, though approximate, is fast and effective for users prioritizing speed. The random approach serves as a baseline and is useful when user-defined constraints apply.
By including these methods, we capture the full spectrum of trade-offs between accuracy and efficiency, addressing optimization objectives for all circuits. The following section details these algorithms and compares their space and time complexities, highlighting their strengths and limitations.

\subsection{Understanding the Algorithms}

\subsubsection{Random Algorithm}

Among the data blocks in Table~\ref{tab:data_block_details}, the `compact' block uses the least space but has the highest time cost for consuming a magic state, while the `fast' block occupies the most space but requires the least time. The `intermediate' block offers a balanced trade-off between space and time.
For the distillation protocols in Table~\ref{tab:distillation_block_details}, the `15-to-1' protocol uses the least space but incurs a high average step count, while the `20-to-4' protocol consumes the most space but minimizes steps. The `116-to-12' protocol provides a balanced approach in terms of space and steps.

The random optimization algorithm assigns a preselected data block and distillation protocol based on the optimization objective. For minimizing tiles, it selects the `compact' data block and `15-to-1' distillation protocol, prioritizing qubit count reduction at the expense of execution time. For balanced optimization, it chooses the `intermediate' data block and `116-to-12' protocol, maintaining a balance between space and execution time. For minimizing steps, it selects the `fast' data block and `20-to-4' protocol, prioritizing reduced execution time at the cost of more qubits. While efficient, this approach does not guarantee globally optimal results.

Mathematically, let $D$ be the set of available data blocks:
$
D = \{\text{compact}, \text{intermediate}, \text{fast}\}
$
and $P$ be the set of available distillation protocols:
$
P = \{15\text{-to-}1, 116\text{-to-}12, 20\text{-to-}4\}
$
with $O$ representing the set of optimization objectives:
$O$ = \{\text{Min Tiles}, \text{Balanced}, \text{Min Steps}\}.
We define a mapping function that assigns a predefined data block and protocol for each optimization type:

\[
f(O) = 
\begin{cases} 
(\text{compact}, \text{15-to-1}), & \text{if } O = \text{Min Tiles} \\
(\text{intermediate}, \text{116-to-12}), & \text{if } O = \text{Balanced} \\
(\text{fast}, \text{20-to-4}), & \text{if } O = \text{Min Steps}
\end{cases}
\]

Thus, the final selected configuration is given by:
$
(d^*, p^*) = f(O)
$,
where $(d^*, p^*)$ represents the assigned data block and distillation protocol for a given optimization objective. 

\subsubsection{Brute Force Algorithm}

Algorithm~\ref{alg:brute_force} systematically explores all possible protocol sequences of length \( 1 \) to \( L \) from the set of available \textit{magic state distillation protocols} \( \mathcal{P} \), optimizing for minimal \textit{tile cost} and \textit{steps}. For each \textit{data block type} \( d \), the required \textit{data tiles} are computed using the relation between the type of data block and the number of qubits in the circuit from Table \ref{tab:data_block_details}. Each candidate protocol $p$ is evaluated by simulating the execution process while tracking the current time step \( s_{\text{current}} \), and available magic states \( M \) keeping track of the stalls in the procedure incurred due to the unavailability of magic states. The algorithm iterates over each \textit{column processing time} \( s_i \) and ensures that a magic state is available before processing by selecting the earliest available protocol (Line 11). If no magic states are available, execution stalls, and more magic states are produced (Line 12). Once a magic state is available, execution proceeds, consuming one magic state per column, and updating the time step. After processing all columns, the \textit{total tile cost} is computed as: \(T_{\text{total}} = T_d + \sum_{p \in \mathcal{P}_s} D_p\), where \( D_p \) represents the \textit{tile cost} of each selected protocol. The resulting configurations, including protocols, final time step, and tile cost, are returned and the minimum tiles and time step along with a balanced intermediate are computed. 

\begin{algorithm}
\caption{Brute Force Algorithm}
\label{alg:brute_force}
\begin{algorithmic}[1]

\Procedure {brute-force}{$n$, $D$, $\mathcal{P}$, $L$}
    \State $\mathcal{P}_s$ $\gets$ all number of protocols from $1$ to $L$ from $\mathcal{P}$
    \State $\mathcal{R} \gets \phi$
    
    \For{each $d$ in $D$}
        \State $T_d \gets \text{TILES}(d,n)$
        \For{each $p$ in $\mathcal{P}_s$}
            \State $s_{\text{current}} \gets 0, M \gets 0$
            \State $s_{\text{next}}(p) \gets \text{STEPS}(p), \forall p \in \mathcal{P}_s$
            
            \For{each column time step $s_i$}
                \While{$M = 0$}
                    \State $p^* = \arg\min_{p \in \mathcal{P}_s} s_{\text{next}}(p)$
                    \State $M \gets M + k_{p^*}$ 
                    \State $s_{\text{next}}(p^*) \gets s_{\text{next}}(p^*) + \text{STEPS}({p^*})$
                \EndWhile
                \State $M \gets M - 1, s_{\text{current}} \gets s_{\text{current}} + s_i$
            \EndFor
            
            \State $T_{\text{total}} = T_d + \sum_{p \in \mathcal{P}_s} D_p$
            \State $\mathcal{R} \gets \mathcal{R} \cup \{(d, \mathcal{P}_s, s_{\text{current}}, T_{\text{total}})\}$
        \EndFor
    \EndFor
    \State \Return $\mathcal{R}$
\EndProcedure
\end{algorithmic}
\end{algorithm}

\subsubsection{Dynamic Programming Algorithm}
The problem of optimizing the distillation protocols for a particular quantum circuit can be broken down into smaller subproblems with overlapping solutions. We define $Cost$ as the cost of consuming a magic state per \textit{column} in the circuit based on the time steps taken to produce and utilize the magic state, and recursively build on top of it to solve for the entire circuit, thus obtaining an optimal substructure for the solution to the subproblems. Algorithm~\ref{alg:dynamic-programming} implements a dynamic programming process to the above mentioned approach, selecting an optimal sequence of magic state distillation protocols $P$ to minimize both \textit{resource cost} and \textit{execution time}. A DP table indexed by column index \( i \) and time step \( s \) is maintained, where each entry stores the \textit{minimum cost} \( C(i, s) \), the number of available \textit{magic states} \( M(i, s) \), and the next \textit{availability time} \( S_{\text{next}, j}(i, s) \) for each protocol \( j \in P\). 

For every protocol $j \in P$; the time steps requires for its completion is set as \(S_{\text{next}, j}(i, s) = \max(s, S_{\text{next}, j}(i-1, s_{\text{prev}})) + S_j\), for each column $i$ and every time step $s_{prev}$; and the magic states are updates using the recursion: \(M(i, t) = M(i-1, s_{\text{prev}}) + k_j\). In case there are not enough magic states for the current time step, the sequence is stalled to let the optimum protocol produce a magic state, followed by decrementing $M(i,s)$ by one to denote the consumption of one magic state. The $Cost$ table is updated by taking into account the number of tiles produced by the protocol and the cost of the previous iteration: 
\[C'(i, s) \gets C'(i-1, s_{\text{prev}}) + \text{TILES}(j,n) + S(i, s)\]
The base cases of the recursion are: \(C(0, 0) =  M(0, 0) = 0\). Based on the optimal cost, the selected protocols and other configurations are returned.

\begin{algorithm}[]
\caption{Dynamic Programming Algorithm}
\label{alg:dynamic-programming}
\begin{algorithmic}[1]
\Procedure{dp}{$n$, $P$, $C$}
    \State $C'(i, s) \gets \infty$, $M(i, s) \gets 0$
    \State $S_{\text{next}, j}(0, 0) \gets 0$ for all $j \in P$
    \State $Cost(0, 0) \gets 0, M(0, 0) \gets 0$
    \State $\mathcal{R} \gets \phi$
    
    \For{$i = 1$ to $C$}
        \For{each $s_{\text{prev}}$ in DP table}
            \For{each $j \in P$}
                \State \raggedright $S_{\text{next}, j}(i, s) = \max(s, S_{\text{next}, j}(i-1, s_{\text{prev}})) + S_j$
                
                \If{$S_{\text{next}, j}(i-1, s_{\text{prev}}) \leq s$}
                    \State $M(i, s) \gets M(i-1, s_{\text{prev}}) + k_j$
                \EndIf
    
                \If{$M(i, s) > 0$}
                    \State $M(i, s) \gets M(i, s) - 1$
                \Else
                    \State Stall
                \EndIf
    
                \State $s \gets s_{\text{prev}} + s_{j, i}$ 
                \State $C'(i, s) \gets C'(i-1, s_{\text{prev}}) + \text{TILES}(j,n) + S(i, s)$
    
                \If{$C'(i, s) < C'_{\text{best}}(i, s)$}
                    \State $C'_{\text{best}}(i, s) \gets C'(i, s)$
                    \State $\mathcal{R} \gets \mathcal{R} \cup \{(j, C'_{\text{best}}(i, s))\}$
                \EndIf
            \EndFor
        \EndFor
    \EndFor
    
    \State \Return $\mathcal{R}$
\EndProcedure
\end{algorithmic}
\end{algorithm}

\subsubsection{Greedy Algorithm}
Algorithm~\ref{alg:greedy_distillation} implements the greedy approach to the problem by selecting an optimal sequence of distillation protocols to minimize both tile resources and execution time steps. It iterates over different data block types \( d \), computing the required data tiles \( T_d \) and initializing time step variables \( s_{\text{current}} \), \( s_{\text{next}} \), and available magic states \( M \). A greedy selection is applied to iteratively choose the best protocol \( p^* \), determined by minimizing the metric: \(M_p = \frac{D_p}{k_p} + S_p\), where \( D_p \) is the distillation tile cost, \( k_p \) is the number of produced magic states, and \( S_p \) is the distillation time. After selecting \( L \) protocols, column execution is processed by consuming one magic state per step while updating \( s_{\text{current}} \); if no magic states are available, execution stalls until the next production cycle. The total tile cost is computed as:
    \(T_{\text{total}} = T_d + \sum_{p \in \mathcal{P}_s} D_p\).
Again, we obtain the optimal protocols and time step configurations from the function call to compute the minimum tiles and the time steps required for the circuit.

\begin{algorithm}
\caption{Greedy Algorithm}
\label{alg:greedy_distillation}
\begin{algorithmic}[1]
\Procedure{Greedy}{$n$, $D$, $L$, $P$}
    \State $\mathcal{R} \gets \phi$, $\mathcal{P}_s \gets \phi$
    \For{each $d$ in $D$}
        \State $T_d \gets \text{TILES}(n)$
        \State $s_{\text{current}} \gets 0, M \gets 0, s_{\text{next}} \gets 0$
        
        \While{$|\mathcal{P}_s| < L$}
            \State $ M_p = \frac{D_p}{k_p} + T_p$
            \State $p^* = \arg\min_{p \in \mathcal{P}} M_p$
            \State $\mathcal{P}_s \gets \mathcal{P}_s \cup p^*$, $M \gets M + k_{p^*}$
            \If{$s_{\text{current}} \geq s_{\text{next}}$}
                \State $s_{\text{next}} \gets s_{\text{current}} + S_{p^*}$
            \Else
                \State $s_{\text{current}} \gets s_{\text{next}}$
            \EndIf
        \EndWhile
    
        \For{each column execution time $s_i$}
            \If{$M = 0$}
                \State $s_{\text{current}} \gets s_{\text{next}}$, $M \gets M + k_{p^*}$
            \EndIf
            \State $M \gets M - 1$, $s_{\text{current}} \gets s_{\text{current}} + s_i$
        \EndFor
    
        \State $T_{\text{total}} = T_d + \sum_{p \in \mathcal{P}_s} D_p$

        \State $\mathcal{R} \gets \mathcal{R} \cup \{(d, \mathcal{P}_s, s_{\text{current}}, T_{\text{total}})\}$
    \EndFor
    \State \Return $\mathcal{R}$
\EndProcedure
\end{algorithmic}
\end{algorithm}

\subsubsection{Finding the Extreme and Balanced results:}

Random algorithm 
does not require separate identification of the `minimize tiles', `minimize steps', and `balanced' results. However, for the remaining three algorithms, Brute Force, Dynamic Programming, and Greedy, we do need to explicitly determine these results. Algorithms~\ref{alg:brute_force},~\ref{alg:dynamic-programming} and~\ref{alg:greedy_distillation} return results \( \mathcal{R} \) which is used as input to calculate the `minimize tiles', `minimize steps', and `balanced' results.

To determine the `minimize tiles' result, we select the configuration with the smallest total number of tiles (\textit{Total Tiles}). If multiple configurations have the same \textit{Total Tiles}, we choose the one with the smallest total computation time (\textit{Total steps}). This ensures that the primary objective of minimizing spatial resources is achieved, while also preferring time-efficient solutions when spatial requirements are identical. 
Prioritizing configurations with lower steps in the case of ties ensures a balance between resource efficiency and execution speed.
To determine the `minimize steps' result, we select the configuration with the smallest \textit{Total Steps}. In cases where multiple configurations have the same \textit{Total Steps}, we choose the one with the smallest \textit{Total Tiles}. This ensures that temporal efficiency is the primary focus, while spatial efficiency serves as a secondary consideration. 

To identify the balanced result, we calculate the midpoint between the `minimize tiles' and `minimize steps' results. Specifically, we compute the average number of tiles and the average computation time as:
\[
Midpoint_T = \frac{\text{Total Tiles (Min Tiles)} + \text{Total Tiles (Min Time)}}{2},
\]
\[
Midpoint_S = \frac{\text{Total Steps (Min Tiles)} + \text{Total Steps (Min Steps)}}{2}.
\]
Next, we evaluate all configurations in \( \mathcal{R} \) to find the one closest to this midpoint using the Euclidean distance:
\[
\text{d} = \sqrt{(\text{Total Tiles} - Midpoint_T)^2 + (\text{Total Steps} - Midpoint_S)^2}.
\]
The configuration with the smallest distance is selected as the balanced result, representing the best trade-off between spatial and temporal efficiency. The balanced result is especially valuable in scenarios where both tiles and time are equally critical, ensuring that neither metric is overly prioritized at the expense of the other. 



\subsection{Time and Space Complexity}

Each algorithm's time and space complexities are explained in detail below, with a summary provided in Table~\ref{tab:algorithm_complexities}. 

\begin{table}[]
\centering
\fontsize{8.5pt}{9.5pt}\selectfont
\caption{Time and Space Complexities of Various Algorithms}
\begin{tabular}{c||cc}
\textbf{Algorithm} & \textbf{Time Complexity} & \textbf{Space Complexity} \\
\hline \hline
Random & $O(1)$ & $O(1)$ \\
BF & $O(3 \cdot 4^L \cdot C \cdot S_{\text{max}})$ & $O(4^L + 3)$ \\
DP & $O(C \cdot S_{\text{max}} \cdot 4)$ & $O(C \cdot S_{\text{max}} + 4)$ \\
Greedy & $O(L \cdot 4 + S_{\text{max}})$ & $O(L + 4)$ \\
\hline \hline
\end{tabular}
\label{tab:algorithm_complexities}
\end{table}

\subsubsection{Random Algorithm:}

Since the random algorithm operates based on a predefined selection process and updates results instantly, its time and space complexity are both $O(1)$.

\subsubsection{Brute-Force Algorithm:}

The brute force algorithm generates all possible numbers of protocols up to \(L = C\), for \(P = 4\). For each sequence, the algorithm evaluates its performance across all columns (\(C\)) and time steps (\(S_{\text{max}}\)). Additionally, it iterates over all unique data block types (\(D = 3\)). This results in the following total time complexity:
$
O(D \cdot 4^L \cdot C \cdot S_{\text{max}})
$.
The space complexity of brute force is dominated by the storage required for all protocol sequences, \(O(4^L)\), and additional data for each block type (\(D = 3\)):
$
O(4^L + D) = O(4^L + 3)
$.
Brute force complexity grows exponentially with \(L\) and scales with both \(C\) and \(S_{\text{max}}\), making it computationally infeasible for larger circuits or sequence lengths.

\subsubsection{Dynamic Programming Algorithm:}

The time complexity of the dynamic programming algorithm arises from iterating through the columns, steps, and protocols. The outer loop iterates over the number of columns (\(C\)), representing the sequential processing of each column in the circuit. Within each column, the algorithm tracks all possible time steps (\(S_{\text{max}}\)), which defines the maximum step granularity. For every combination of a column and steps, the algorithm iterates through all possible protocols (\(P = 4\)). As a result, the total time complexity is:
$
O(C \cdot S_{\text{max}} \cdot P) = O(C \cdot S_{\text{max}} \cdot 4)
$.
The space complexity of dynamic programming is driven by the size of the DP table, which tracks costs for all combinations of columns and steps. This requires \(O(C \cdot S_{\text{max}})\) storage. Additionally, the protocol parameters occupy \(O(P)\) space. Therefore, the total space complexity is:
$
O(C \cdot S_{\text{max}} + P) = O(C \cdot S_{\text{max}} + 4)
$.
Dynamic programming scales polynomially with the circuit depth and step granularity, making it more efficient than brute force but still computationally intensive for large \(C\) or \(S_{\text{max}}\).

\subsubsection{Greedy Algorithm:}

The greedy algorithm operates by selecting the best protocol for each sequence position. It iterates through all unique data block types (\(D = 3\)), representing the three types of input configurations. For each data block type, it builds protocol sequences up to a maximum length (\(L = C\)). Within this loop, the algorithm evaluates \(P = 4\) protocols at each step to select the one minimizing a specific cost function. This leads to the following total time complexity:
$
O(D \cdot L \cdot P) = O(3 \cdot L \cdot 4) = O(L \cdot 4)
$
When considering step stall computations, \(S_{\text{max}}\) is added to the complexity:
$
O(L \cdot 4 + S_{\text{max}})
$.
The space complexity of the greedy algorithm is determined by the protocol sequence storage, which is \(O(L)\), and the protocol parameters, \(O(P)\). Combining these terms, the total space complexity is:
$
O(L + P) = O(L + 4)
$.
The greedy algorithm remains highly efficient, with both time and space complexity scaling minimally with the circuit size and sequence length.

\subsubsection{Comparing the Algorithms:}

\begin{figure}
    \centering
    \includegraphics[width=1\linewidth]{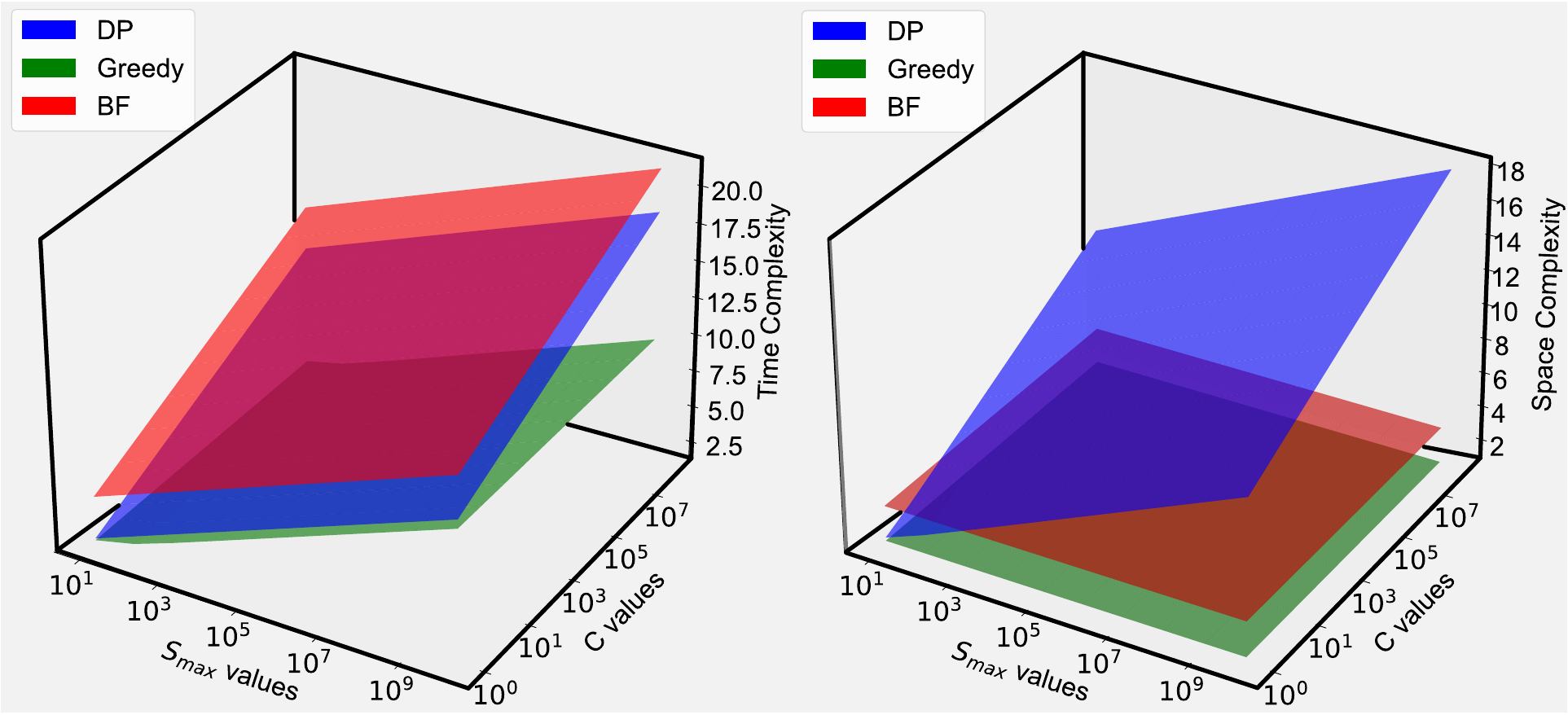}
    \caption{Computational Complexity Analysis of Optimization Algorithms: 
    The left figure shows time step complexity and the right figure shows space complexity for the proposed algorithms, evaluating their scalability and suitability for different problem sizes. The analysis includes up to $10^7$ columns, $10^5$ qubits, and $10^{13}$ T gates.
}
    \label{fig:algorithm_complexity}
\end{figure}

Fig.~\ref{fig:algorithm_complexity} compares the time step complexity (left) and space complexity (right) for the proposed algorithms.
To analyze the complexities of the algorithms, the circuit property range is extended to include up to $10^7$ columns, corresponding to a maximum of $10^5$ qubits and $10^{13}$ T gates.

The brute force algorithm consistently provides the best results because it evaluates every possible protocol sequence exhaustively. 
However, this thoroughness comes with a significant computational cost, as the time complexity of brute force grows exponentially with columns, making it the most time-intensive algorithm. In comparison, dynamic programming and greedy algorithms are far more efficient, with dynamic programming scaling polynomially and the greedy algorithm scaling linearly in time complexity. In terms of space complexity, dynamic programming requires significantly more storage (than brute force and greedy algorithms) due to a large table that is used to store intermediate results for every combination of columns and steps, leading to space complexity that scales. 
Overall, brute force delivers the best results at the cost of the highest time complexity, dynamic programming strikes a balance but requires significant storage, and the greedy algorithm is the most computationally efficient, albeit with suboptimal results. These trade-offs make the choice of algorithm dependent on the problem size and available resources.


\section{Comparison and Evaluation} \label{sec:evaluation}

\begin{table*}[]
\centering
\caption{Results of Algorithmic and Optimization Strategies across Various Circuit Categories and Subcategories}
\begin{adjustbox}{angle=90}
\fontsize{9pt}{10.2pt}\selectfont
\begin{tabular}{ccc||ccc||cc||ccccc}
\hline \hline
\multicolumn{3}{c||}{\textbf{Classification}}                                                                                                                                             & \multicolumn{3}{c||}{\textbf{Original Circuit}}                                                                   & \multicolumn{2}{c||}{\textbf{Algorithm}}                              & \multicolumn{5}{c}{\textbf{Optimized Results}}                                                                             \\ \hline
\multicolumn{1}{c|}{\textbf{Category}}                                                                   & \multicolumn{1}{c|}{\textbf{Sub Cat.}}                 & \textbf{Class}       & \multicolumn{1}{c|}{\textbf{\#Q}}         & \multicolumn{1}{c|}{\textbf{\#C}}          & \textbf{\#T}            & \multicolumn{1}{c|}{\textbf{Type}}           & \textbf{Optimization} & \textbf{Data Block} & \textbf{Protocols}                           & \textbf{\#Steps} & \textbf{\#Idle} & \textbf{\#Tiles} \\ \hline \hline
\multicolumn{1}{c|}{\multirow{16}{*}{\textbf{\begin{tabular}[c]{@{}c@{}}Circuit\\ Depth\end{tabular}}}}  & \multicolumn{1}{c|}{\multirow{8}{*}{\textbf{Shallow}}} & \multirow{8}{*}{SHS} & \multicolumn{1}{c|}{\multirow{8}{*}{10}}  & \multicolumn{1}{c|}{\multirow{8}{*}{1}}    & \multirow{8}{*}{10}     & \multicolumn{1}{c|}{\multirow{2}{*}{Random}} & Min Tiles             & Compact             & 15-to-1                                      & 19               & 11              & 31               \\
\multicolumn{1}{c|}{}                                                                                    & \multicolumn{1}{c|}{}                                  &                      & \multicolumn{1}{c|}{}                     & \multicolumn{1}{c|}{}                      &                         & \multicolumn{1}{c|}{}                        & Min Steps             & Fast                & 20-to-4                                      & 18               & 17              & 34               \\ \cline{7-13} 
\multicolumn{1}{c|}{}                                                                                    & \multicolumn{1}{c|}{}                                  &                      & \multicolumn{1}{c|}{}                     & \multicolumn{1}{c|}{}                      &                         & \multicolumn{1}{c|}{\multirow{2}{*}{BF}}     & Min Tiles             & compact             & 15-to-1                                      & 11               & 3               & 29               \\
\multicolumn{1}{c|}{}                                                                                    & \multicolumn{1}{c|}{}                                  &                      & \multicolumn{1}{c|}{}                     & \multicolumn{1}{c|}{}                      &                         & \multicolumn{1}{c|}{}                        & Min Steps             & fast                & 15-to-1 ; 225-to-1                           & 11               & 10              & 216              \\ \cline{7-13} 
\multicolumn{1}{c|}{}                                                                                    & \multicolumn{1}{c|}{}                                  &                      & \multicolumn{1}{c|}{}                     & \multicolumn{1}{c|}{}                      &                         & \multicolumn{1}{c|}{\multirow{2}{*}{DP}}     & Min Tiles             & compact             & 15-to-1                                      & 11               & 3               & 29               \\
\multicolumn{1}{c|}{}                                                                                    & \multicolumn{1}{c|}{}                                  &                      & \multicolumn{1}{c|}{}                     & \multicolumn{1}{c|}{}                      &                         & \multicolumn{1}{c|}{}                        & Min Steps             & fast                & 15-to-1 ; 20-to-4                            & 11               & 10              & 54               \\ \cline{7-13} 
\multicolumn{1}{c|}{}                                                                                    & \multicolumn{1}{c|}{}                                  &                      & \multicolumn{1}{c|}{}                     & \multicolumn{1}{c|}{}                      &                         & \multicolumn{1}{c|}{\multirow{2}{*}{Greedy}} & Min Tiles             & compact             & 20-to-4 (2)                                  & 25               & 17              & 46               \\
\multicolumn{1}{c|}{}                                                                                    & \multicolumn{1}{c|}{}                                  &                      & \multicolumn{1}{c|}{}                     & \multicolumn{1}{c|}{}                      &                         & \multicolumn{1}{c|}{}                        & Min Steps             & fast                & 20-to-4 (2)                                  & 18               & 17              & 57               \\ \cline{2-13}
\multicolumn{1}{c|}{}                                                                                    & \multicolumn{1}{c|}{\multirow{8}{*}{\textbf{Deep}}}    & \multirow{8}{*}{DLL} & \multicolumn{1}{c|}{\multirow{8}{*}{100}} & \multicolumn{1}{c|}{\multirow{8}{*}{5833}} & \multirow{8}{*}{5833}   & \multicolumn{1}{c|}{\multirow{2}{*}{Random}} & Min Tiles             & Compact             & 15-to-1                                      & 64171            & 40550           & 211              \\
\multicolumn{1}{c|}{}                                                                                    & \multicolumn{1}{c|}{}                                  &                      & \multicolumn{1}{c|}{}                     & \multicolumn{1}{c|}{}                      &                         & \multicolumn{1}{c|}{}                        & Min Steps             & Fast                & 20-to-4                                      & 24804            & 21085           & 214              \\ \cline{7-13} 
\multicolumn{1}{c|}{}                                                                                    & \multicolumn{1}{c|}{}                                  &                      & \multicolumn{1}{c|}{}                     & \multicolumn{1}{c|}{}                      &                         & \multicolumn{1}{c|}{\multirow{2}{*}{BF}}     & Min Tiles             & compact             & 15-to-1                                      & 40524            & 17247           & 164              \\
\multicolumn{1}{c|}{}                                                                                    & \multicolumn{1}{c|}{}                                  &                      & \multicolumn{1}{c|}{}                     & \multicolumn{1}{c|}{}                      &                         & \multicolumn{1}{c|}{}                        & Min Steps             & fast                & 15-to-1 ; 116-to-12 ; 225-to-1 (2);  20-to-4 & 7185             & 3501            & 649              \\ \cline{7-13} 
\multicolumn{1}{c|}{}                                                                                    & \multicolumn{1}{c|}{}                                  &                      & \multicolumn{1}{c|}{}                     & \multicolumn{1}{c|}{}                      &                         & \multicolumn{1}{c|}{\multirow{2}{*}{DP}}     & Min Tiles             & compact             & 15-to-1                                      & 40656            & 17133           & 164              \\
\multicolumn{1}{c|}{}                                                                                    & \multicolumn{1}{c|}{}                                  &                      & \multicolumn{1}{c|}{}                     & \multicolumn{1}{c|}{}                      &                         & \multicolumn{1}{c|}{}                        & Min Steps             & fast                & 15-to-1 ;  20-to-4 (2)                       & 11340            & 7644            & 267              \\ \cline{7-13} 
\multicolumn{1}{c|}{}                                                                                    & \multicolumn{1}{c|}{}                                  &                      & \multicolumn{1}{c|}{}                     & \multicolumn{1}{c|}{}                      &                         & \multicolumn{1}{c|}{\multirow{2}{*}{Greedy}} & Min Tiles             & compact             & 20-to-4 (2)                                  & 23245            & 17              & 181              \\
\multicolumn{1}{c|}{}                                                                                    & \multicolumn{1}{c|}{}                                  &                      & \multicolumn{1}{c|}{}                     & \multicolumn{1}{c|}{}                      &                         & \multicolumn{1}{c|}{}                        & Min Steps             & fast                & 20-to-4 (2)                                  & 11324            & 7645            & 256              \\ \hline \hline
\multicolumn{1}{c|}{\multirow{16}{*}{\textbf{\begin{tabular}[c]{@{}c@{}}T Gate\\ Density\end{tabular}}}} & \multicolumn{1}{c|}{\multirow{8}{*}{\textbf{Low}}}     & \multirow{8}{*}{SLS} & \multicolumn{1}{c|}{\multirow{8}{*}{30}}  & \multicolumn{1}{c|}{\multirow{8}{*}{625}}  & \multirow{8}{*}{4401}   & \multicolumn{1}{c|}{\multirow{2}{*}{Random}} & Min Tiles             & Compact             & 15-to-1                                      & 6883             & 2015            & 71               \\
\multicolumn{1}{c|}{}                                                                                    & \multicolumn{1}{c|}{}                                  &                      & \multicolumn{1}{c|}{}                     & \multicolumn{1}{c|}{}                      &                         & \multicolumn{1}{c|}{}                        & Min Steps             & Fast                & 20-to-4                                      & 2670             & 2045            & 74               \\ \cline{7-13} 
\multicolumn{1}{c|}{}                                                                                    & \multicolumn{1}{c|}{}                                  &                      & \multicolumn{1}{c|}{}                     & \multicolumn{1}{c|}{}                      &                         & \multicolumn{1}{c|}{\multirow{2}{*}{BF}}     & Min Tiles             & compact             & 15-to-1                                      & 6875             & 2043            & 59               \\
\multicolumn{1}{c|}{}                                                                                    & \multicolumn{1}{c|}{}                                  &                      & \multicolumn{1}{c|}{}                     & \multicolumn{1}{c|}{}                      &                         & \multicolumn{1}{c|}{}                        & Min Steps             & fast                & 20-to-4 (2)                                  & 1201             & 576             & 103              \\ \cline{7-13} 
\multicolumn{1}{c|}{}                                                                                    & \multicolumn{1}{c|}{}                                  &                      & \multicolumn{1}{c|}{}                     & \multicolumn{1}{c|}{}                      &                         & \multicolumn{1}{c|}{\multirow{2}{*}{DP}}     & Min Tiles             & compact             & 15-to-1                                      & 6875             & 2055            & 59               \\
\multicolumn{1}{c|}{}                                                                                    & \multicolumn{1}{c|}{}                                  &                      & \multicolumn{1}{c|}{}                     & \multicolumn{1}{c|}{}                      &                         & \multicolumn{1}{c|}{}                        & Min Steps             & fast                & 15-to-1 ;  20-to-4 (2)                       & 1923             & 1298            & 114              \\ \cline{7-13} 
\multicolumn{1}{c|}{}                                                                                    & \multicolumn{1}{c|}{}                                  &                      & \multicolumn{1}{c|}{}                     & \multicolumn{1}{c|}{}                      &                         & \multicolumn{1}{c|}{\multirow{2}{*}{Greedy}} & Min Tiles             & compact             & 20-to-4 (2)                                  & 4855             & 17              & 76               \\
\multicolumn{1}{c|}{}                                                                                    & \multicolumn{1}{c|}{}                                  &                      & \multicolumn{1}{c|}{}                     & \multicolumn{1}{c|}{}                      &                         & \multicolumn{1}{c|}{}                        & Min Steps             & fast                & 20-to-4 (2)                                  & 1201             & 576             & 103              \\ \cline{2-13} 
\multicolumn{1}{c|}{}                                                                                    & \multicolumn{1}{c|}{\multirow{8}{*}{\textbf{High}}}    & \multirow{8}{*}{SHM} & \multicolumn{1}{c|}{\multirow{8}{*}{50}}  & \multicolumn{1}{c|}{\multirow{8}{*}{417}}  & \multirow{8}{*}{15741}  & \multicolumn{1}{c|}{\multirow{2}{*}{Random}} & Min Tiles             & Compact             & 15-to-1                                      & 4595             & 1259            & 111              \\
\multicolumn{1}{c|}{}                                                                                    & \multicolumn{1}{c|}{}                                  &                      & \multicolumn{1}{c|}{}                     & \multicolumn{1}{c|}{}                      &                         & \multicolumn{1}{c|}{}                        & Min Steps             & Fast                & 20-to-4                                      & 1786             & 1369            & 114              \\ \cline{7-13} 
\multicolumn{1}{c|}{}                                                                                    & \multicolumn{1}{c|}{}                                  &                      & \multicolumn{1}{c|}{}                     & \multicolumn{1}{c|}{}                      &                         & \multicolumn{1}{c|}{\multirow{2}{*}{BF}}     & Min Tiles             & compact             & 15-to-1                                      & 4587             & 1251            & 89               \\
\multicolumn{1}{c|}{}                                                                                    & \multicolumn{1}{c|}{}                                  &                      & \multicolumn{1}{c|}{}                     & \multicolumn{1}{c|}{}                      &                         & \multicolumn{1}{c|}{}                        & Min Steps             & fast                & 20-to-4 (2)                                  & 800              & 383             & 148              \\ \cline{7-13} 
\multicolumn{1}{c|}{}                                                                                    & \multicolumn{1}{c|}{}                                  &                      & \multicolumn{1}{c|}{}                     & \multicolumn{1}{c|}{}                      &                         & \multicolumn{1}{c|}{\multirow{2}{*}{DP}}     & Min Tiles             & compact             & 15-to-1                                      & 4587             & 1251            & 89               \\
\multicolumn{1}{c|}{}                                                                                    & \multicolumn{1}{c|}{}                                  &                      & \multicolumn{1}{c|}{}                     & \multicolumn{1}{c|}{}                      &                         & \multicolumn{1}{c|}{}                        & Min Steps             & fast                & 15-to-1 ;  20-to-4 (2)                       & 1287             & 870             & 159              \\ \cline{8-13} 
\multicolumn{1}{c|}{}                                                                                    & \multicolumn{1}{c|}{}                                  &                      & \multicolumn{1}{c|}{}                     & \multicolumn{1}{c|}{}                      &                         & \multicolumn{1}{c|}{\multirow{2}{*}{Greedy}} & Min Tiles             & compact             & 20-to-4 (2)                                  & 3353             & 17              & 106              \\
\multicolumn{1}{c|}{}                                                                                    & \multicolumn{1}{c|}{}                                  &                      & \multicolumn{1}{c|}{}                     & \multicolumn{1}{c|}{}                      &                         & \multicolumn{1}{c|}{}                        & Min Steps             & fast                & 20-to-4 (2)                                  & 800              & 383             & 148              \\ \hline \hline
\multicolumn{1}{c|}{\multirow{16}{*}{\textbf{\begin{tabular}[c]{@{}c@{}}Qubit\\ Count\end{tabular}}}}    & \multicolumn{1}{c|}{\multirow{8}{*}{\textbf{Small}}}   & \multirow{8}{*}{MHS} & \multicolumn{1}{c|}{\multirow{8}{*}{20}}  & \multicolumn{1}{c|}{\multirow{8}{*}{1333}} & \multirow{8}{*}{21383}  & \multicolumn{1}{c|}{\multirow{2}{*}{Random}} & Min Tiles             & Compact             & 15-to-1                                      & 14671            & 4003            & 51               \\
\multicolumn{1}{c|}{}                                                                                    & \multicolumn{1}{c|}{}                                  &                      & \multicolumn{1}{c|}{}                     & \multicolumn{1}{c|}{}                      &                         & \multicolumn{1}{c|}{}                        & Min Steps             & Fast                & 20-to-4                                      & 5679             & 4346            & 54               \\ \cline{7-13} 
\multicolumn{1}{c|}{}                                                                                    & \multicolumn{1}{c|}{}                                  &                      & \multicolumn{1}{c|}{}                     & \multicolumn{1}{c|}{}                      &                         & \multicolumn{1}{c|}{\multirow{2}{*}{BF}}     & Min Tiles             & compact             & 15-to-1                                      & 14663            & 3991            & 44               \\
\multicolumn{1}{c|}{}                                                                                    & \multicolumn{1}{c|}{}                                  &                      & \multicolumn{1}{c|}{}                     & \multicolumn{1}{c|}{}                      &                         & \multicolumn{1}{c|}{}                        & Min Steps             & fast                & 20-to-4 (2)                                  & 2581             & 1248            & 80               \\ \cline{7-13} 
\multicolumn{1}{c|}{}                                                                                    & \multicolumn{1}{c|}{}                                  &                      & \multicolumn{1}{c|}{}                     & \multicolumn{1}{c|}{}                      &                         & \multicolumn{1}{c|}{\multirow{2}{*}{DP}}     & Min Tiles             & compact             & 15-to-1                                      & 14663            & 3997            & 44               \\
\multicolumn{1}{c|}{}                                                                                    & \multicolumn{1}{c|}{}                                  &                      & \multicolumn{1}{c|}{}                     & \multicolumn{1}{c|}{}                      &                         & \multicolumn{1}{c|}{}                        & Min Steps             & fast                & 15-to-1 ;  20-to-4 (2)                       & 4097             & 2764            & 91               \\ \cline{7-13} 
\multicolumn{1}{c|}{}                                                                                    & \multicolumn{1}{c|}{}                                  &                      & \multicolumn{1}{c|}{}                     & \multicolumn{1}{c|}{}                      &                         & \multicolumn{1}{c|}{\multirow{2}{*}{Greedy}} & Min Tiles             & compact             & 20-to-4 (2)                                  & 10684            & 17              & 61               \\
\multicolumn{1}{c|}{}                                                                                    & \multicolumn{1}{c|}{}                                  &                      & \multicolumn{1}{c|}{}                     & \multicolumn{1}{c|}{}                      &                         & \multicolumn{1}{c|}{}                        & Min Steps             & fast                & 20-to-4 (2)                                  & 2581             & 1248            & 80               \\ \cline{2-13} 
\multicolumn{1}{c|}{}                                                                                    & \multicolumn{1}{c|}{\multirow{8}{*}{\textbf{Large}}}   & \multirow{8}{*}{MML} & \multicolumn{1}{c|}{\multirow{8}{*}{90}}  & \multicolumn{1}{c|}{\multirow{8}{*}{3000}} & \multirow{8}{*}{103125} & \multicolumn{1}{c|}{\multirow{2}{*}{Random}} & Min Tiles             & Compact             & 15-to-1                                      & 33008            & 9008            & 191              \\
\multicolumn{1}{c|}{}                                                                                    & \multicolumn{1}{c|}{}                                  &                      & \multicolumn{1}{c|}{}                     & \multicolumn{1}{c|}{}                      &                         & \multicolumn{1}{c|}{}                        & Min Steps             & Fast                & 20-to-4                                      & 12754            & 9754            & 194              \\ \cline{7-13} 
\multicolumn{1}{c|}{}                                                                                    & \multicolumn{1}{c|}{}                                  &                      & \multicolumn{1}{c|}{}                     & \multicolumn{1}{c|}{}                      &                         & \multicolumn{1}{c|}{\multirow{2}{*}{BF}}     & Min Tiles             & compact             & 15-to-1                                      & 33000            & 9000            & 149              \\
\multicolumn{1}{c|}{}                                                                                    & \multicolumn{1}{c|}{}                                  &                      & \multicolumn{1}{c|}{}                     & \multicolumn{1}{c|}{}                      &                         & \multicolumn{1}{c|}{}                        & Min Steps             & fast                & 20-to-4 (2)                                  & 5831             & 2831            & 234              \\ \cline{7-13} 
\multicolumn{1}{c|}{}                                                                                    & \multicolumn{1}{c|}{}                                  &                      & \multicolumn{1}{c|}{}                     & \multicolumn{1}{c|}{}                      &                         & \multicolumn{1}{c|}{\multirow{2}{*}{DP}}     & Min Tiles             & compact             & 15-to-1                                      & 33000            & 9000            & 149              \\
\multicolumn{1}{c|}{}                                                                                    & \multicolumn{1}{c|}{}                                  &                      & \multicolumn{1}{c|}{}                     & \multicolumn{1}{c|}{}                      &                         & \multicolumn{1}{c|}{}                        & Min Steps             & fast                & 15-to-1 ;  20-to-4 (2)                       & 9200             & 6200            & 245              \\ \cline{7-13} 
\multicolumn{1}{c|}{}                                                                                    & \multicolumn{1}{c|}{}                                  &                      & \multicolumn{1}{c|}{}                     & \multicolumn{1}{c|}{}                      &                         & \multicolumn{1}{c|}{\multirow{2}{*}{Greedy}} & Min Tiles             & compact             & 20-to-4 (2)                                  & 24017            & 17              & 166              \\
\multicolumn{1}{c|}{}                                                                                    & \multicolumn{1}{c|}{}                                  &                      & \multicolumn{1}{c|}{}                     & \multicolumn{1}{c|}{}                      &                         & \multicolumn{1}{c|}{}                        & Min Steps             & fast                & 20-to-4 (2)                                  & 5831             & 2831            & 234              \\ \hline \hline
\end{tabular}
\end{adjustbox}
\label{tab:big_table}
\end{table*}

Table~\ref{tab:big_table} presents results from each algorithm, showing the specific columns they produce. Circuits are sampled from each subcategory of the three major circuit categories, and for each case, four algorithms are evaluated under two optimization objectives: minimizing steps and minimizing tiles. The original circuit parameters include the number of qubits, columns, and T gates. The optimization results provide the data block types and distillation protocols used, along with key performance metrics: total execution steps, tiles required, and idle steps. The number of distinct distillation protocols used ranges from 1 to 5. Due to space limitations, the medium subcategory and balanced optimization are excluded from these examples. The table showcases six circuits, evaluated by four algorithms and two optimization objectives, totaling 48 rows of data.

\subsection{Comparing the Optimization Methods}

\begin{figure}
    \centering
    \includegraphics[width=1\linewidth]{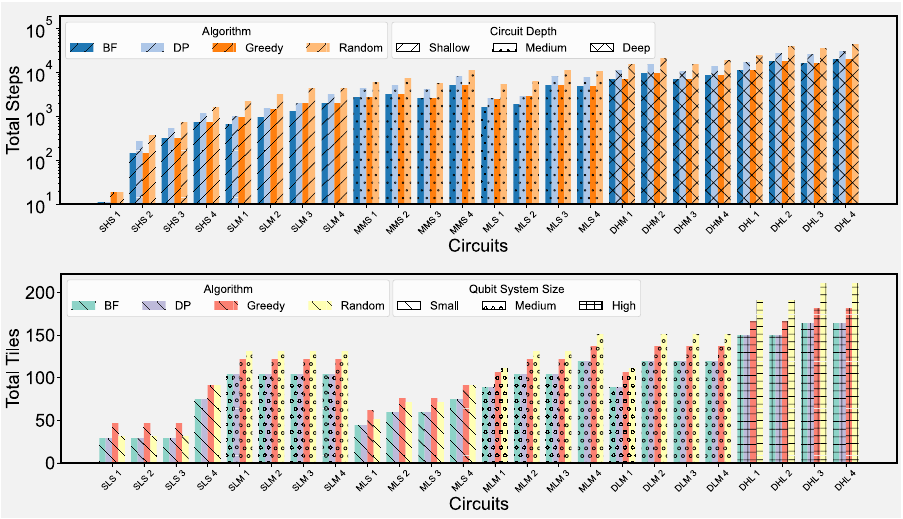}
    \caption{Step and Tile Analysis Across Circuit Categories: 
    (Left) Step counts for six circuit categories, with four algorithms optimized for minimizing tiles. The greedy algorithm performs similarly to brute force due to local step optimization, with step counts increasing with circuit depth. (Right) Tile counts for the same categories, with bar patterns indicating qubit numbers. Dynamic programming outperforms other algorithms, matching brute force results by efficiently reusing solutions. Unlike steps, tile count increases with qubit numbers as larger systems require more resources.
}
    \label{fig:steps_and_tiles_wrt_circuits}
\end{figure}

From the analysis, we note that brute force consistently yields the lowest values in both cases. To evaluate other algorithms, we compare their results against brute force to assess their accuracy.
Figure~\ref{fig:steps_and_tiles_wrt_circuits} (top) presents a plot of total steps for six representative circuit categories, each containing four example circuits. The plot shows total step counts across four algorithms aiming to minimize tiles, with bar patterns indicating circuit depth. The greedy algorithm performs similarly to brute force in minimizing steps. In cases where the circuit structure allows straightforward minimization, such as shallow circuits or those with minimal dependencies, the greedy algorithm’s locally optimal choices often align with the globally optimal solution, leading to similar step counts. Additionally, the total number of steps increases with circuit depth, confirming the impact of structural complexity on step count.

Figure~\ref{fig:steps_and_tiles_wrt_circuits} (bottom) shows the total tile count for six circuit categories, each with four example circuits. Bar patterns represent the number of qubits, ranging from small to large. The plot displays total step counts across four algorithms aiming to minimize steps. The dynamic programming algorithm is the most effective, consistently matching the brute force tile count across all cases. It avoids redundant computations by storing and reusing previously calculated states, ensuring efficiency while evaluating all possible solutions. Unlike step count, which depends on circuit depth, total tile count is mainly influenced by qubit count, increasing with more qubits. Neither tiles nor steps show a direct correlation with T-gate density, as the number of required magic states depends on columns, not T gates.
 
For the optimization objective of minimizing tiles, circuits were analyzed to compare the total tile count of each algorithm against brute force. The percentage increase in tile usage was calculated as \(\text{\% Increase (Tiles)} = \frac{\text{Total Tiles (Alg.)} - \text{Total Tiles (BF)}}{\text{Total Tiles (BF)}} \times 100\). These increases were averaged based on circuit parameters (qubits, columns, and T gates) to assess algorithm performance. For minimizing time, the percentage increase in steps was calculated similarly: \(\text{\% Increase (Steps)} = \frac{\text{Total Steps (Alg.)} - \text{Total Time (BF)}}{\text{Total Time (BF)}} \times 100\).
The results were averaged to evaluate each algorithm’s efficiency in execution time. Table~\ref{tab:percentage_increase} presents the average percentage increase relative to brute force for different optimization objectives. For minimizing tiles, dynamic programming matches brute force with a $0\%$ increase, followed by the random algorithm with approximately a $15\%$ increase. For minimizing steps, the Greedy algorithm is closest to brute force, with a $7\%$ increase.

\begin{table}[]
\centering
\fontsize{8.5pt}{9.5pt}\selectfont
\caption{Average Percentage Increase relative to Brute Force for Different Algorithms under Optimization Objectives: Minimizing Steps and Minimizing Tiles}
\begin{tabular}{c||cc}
\textbf{Algorithm} & \textbf{Steps (Avg. \% Incr.)} & \textbf{Tiles (Avg. \% Incr.)} \\ \hline \hline
\textbf{DP}        & 64.62                           & 0.00                             \\
\textbf{Greedy}    & 7.59                            & 41.33                            \\
\textbf{Random}    & 148.47                          & 14.69                            \\ \hline \hline
\end{tabular}
\label{tab:percentage_increase}
\end{table}

\subsection{Finding the Balance between Tiles and Steps}

Balanced optimization is determined by finding the midpoint between the tile and step values of the extreme points for minimizing tiles and minimizing steps within the algorithm’s search space. While the formula for this balanced point is consistent across algorithms, the resulting optimizations differ due to each algorithm's unique search space. As a result, balanced optimizations often involve different protocols, and brute force may not always provide the most optimal solution.

\begin{figure}
    \centering
    \includegraphics[width=1\linewidth]{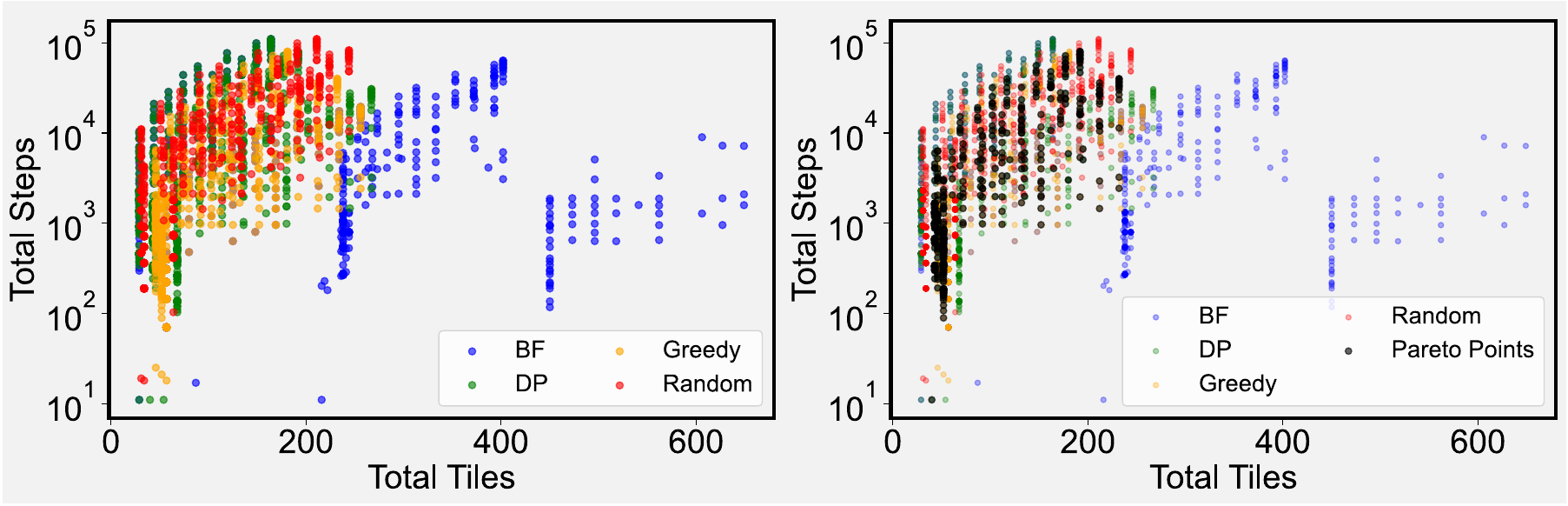}
    \caption{Analysis of Balanced Points and Pareto-Optimal Solutions: 
    (Left) Total steps versus tiles for all circuits and algorithms under three optimization types. (Right) The same data with Pareto-optimal points highlighted, showing their distribution across multiple algorithms.
}
    \label{fig:balanced_with_pareto_points}
\end{figure}

To identify the balanced points, we analyze data from all four algorithms across all circuits and the three optimization types: minimizing tiles, minimizing steps, and balancing. For each case, we plot the total number of steps against the total number of tiles, as shown in Fig.~\ref{fig:balanced_with_pareto_points} (left). This analysis results in 12 points per circuit, corresponding to the three optimization types and four algorithms. Among them, we identify the most optimal point for each circuit using the Pareto algorithm to find the Pareto front. A Pareto front, in the context of this analysis, represents the optimal trade-off between minimizing tiles and minimizing steps. For a given circuit, a solution is considered a Pareto front if no other solution exists that improves one objective (e.g., fewer tiles) without worsening the other (e.g., more steps). Essentially, it is a combination of two points where any further improvement in one metric would lead to a compromise in the other. The Pareto front for each circuit is computed by considering all configurations generated by the four algorithms (brute force, DP, Greedy, Random) across the three optimization types (minimizing tiles, minimizing steps, and balanced). The total tiles and steps for each configuration are compared, and the Pareto front is identified as the unique solution that achieves this optimal balance. For every circuit, there is exactly one Pareto front consisting of a Pareto tile point and a Pareto step point. The scatter plot in Fig.\ref{fig:balanced_with_pareto_points} (right) presents the same data as Fig.\ref{fig:balanced_with_pareto_points} (left), displaying points from all algorithms. However, it additionally highlights the single Pareto for each circuit. This visualization demonstrates that the Pareto-optimal solutions are distributed across multiple algorithms rather than being confined to a single one. 

\begin{figure}
    \centering
    \includegraphics[width=1\linewidth]{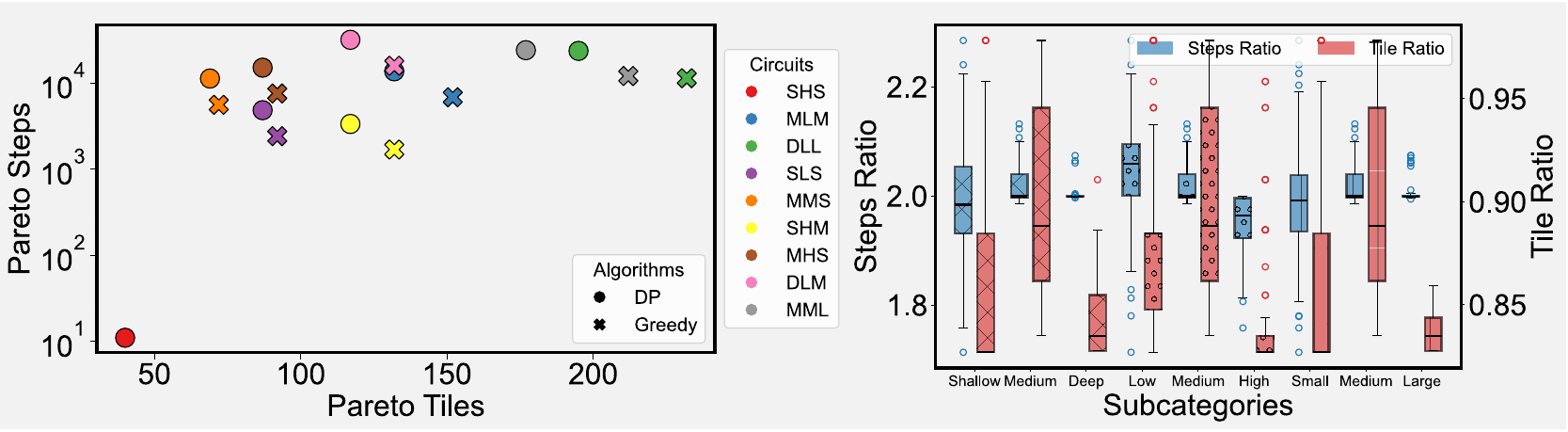}
    \caption{Pareto Fronts and Performance Ratios for DP and Greedy Algorithms: 
    (Left) Nine sample circuits with Pareto fronts showing the trade-off between steps and tiles for DP and Greedy algorithms, with two distinct Pareto fronts for most circuits. (Right) Step and tile ratios (DP/Greedy) across circuit subcategories, where step ratios are always $>1$ and tile ratios $<1$, indicating DP minimizes tiles and Greedy minimizes steps.
}
    \label{fig:pareto_DP_Greedy}
\end{figure}

We observe that only $0.39\%$ of the circuits possess a single Pareto front, while the remaining exhibit two distinct Pareto fronts. The data reveals that these two Pareto fronts consistently correspond to the balanced optimization achieved through the dynamic programming algorithm and the balanced optimization attained using the greedy algorithm. Notably, for circuits with a single Pareto front, it is exclusively the balanced optimization via the dynamic programming algorithm. 

Fig.~\ref{fig:pareto_DP_Greedy} (left) illustrates nine sample circuits, each selected from a different circuit class. For each circuit, we depict their Pareto fronts, specifically showcasing the Pareto steps with respect to the Pareto tiles. Most circuits exhibit Pareto fronts generated by both dynamic programming (DP) and greedy algorithms. In this context, having two Pareto fronts for a circuit signifies the presence of two distinct configurations of optimization outcomes that are both Pareto-optimal. These configurations are not dominated by one another — one may excel in minimizing total steps, while the other achieves better results in minimizing total tiles. Importantly, neither configuration is universally superior across all metrics.
Fig.~\ref{fig:pareto_DP_Greedy} (right) illustrates the step ratio (Pareto steps of DP/Greedy) and tile ratio (Pareto tiles of DP/Greedy) for each case, categorized by circuit classification subcategories. The first set of subcategories corresponds to circuit depth (shallow, medium, deep), the second set to T-gate density (low, medium, high), and the final set to qubit count (small, medium, large). While the ratios do not exhibit any significant variation across these subcategories, the step ratio is consistently greater than 1, and the tile ratio is consistently less than 1. This indicates that, in $100\%$ of cases, the Pareto fronts generated by the DP algorithm involve a higher total number of steps compared to those generated by the Greedy algorithm. Conversely, the Pareto fronts generated by the Greedy algorithm always involve a higher total number of tiles than those produced by the DP algorithm. Thus, while both balanced optimizations — DP and Greedy — offer configurations that are Pareto-optimal, the DP algorithm excels in minimizing the total number of tiles, whereas the Greedy algorithm excels in minimizing the total number of steps.

\subsection{Finding A Generalized Heuristic}

\begin{table*}[]
\centering
\fontsize{8.5pt}{9.5pt}\selectfont
\caption{Heuristic Guide for Selecting Optimization Strategies Across Algorithms Based on Optimization Objectives \\ (Legend: True Results = \ding{51}, Nearly Exact Results = \ding{72}, Close Estimate = \ding{109}, Not Recommended = \ding{55} )}
\begin{tabular}{cc||cccccccccccc}
\hline \hline
\multicolumn{2}{c||}{\multirow{3}{*}{\textbf{Objective}}}                       & \multicolumn{12}{c}{\textbf{Algorithms and Optimization Strategies}}                                                                                                                                                                                                         \\ \cline{3-14} 
\multicolumn{2}{c||}{}                                                        & \multicolumn{3}{c|}{\textbf{Brute Force}}                              & \multicolumn{3}{c|}{\textbf{DP}}                                       & \multicolumn{3}{c|}{\textbf{Greedy}}                                   & \multicolumn{3}{c}{\textbf{Random}}               \\ \cline{3-14} 
\multicolumn{2}{c||}{}                                                        & \textbf{Min. S} & \textbf{Min. T} & \multicolumn{1}{c|}{\textbf{Bal.}} & \textbf{Min. S} & \textbf{Min. T} & \multicolumn{1}{c|}{\textbf{Bal.}} & \textbf{Min. S} & \textbf{Min. T} & \multicolumn{1}{c|}{\textbf{Bal.}} & \textbf{Min. S} & \textbf{Min. T} & \textbf{Bal.} \\ \hline \hline
\multicolumn{2}{c||}{\textbf{Minimize Steps}}                                     & \ding{51}               & \ding{55}               & \multicolumn{1}{c|}{\ding{55}}             & \ding{55}               & \ding{55}               & \multicolumn{1}{c|}{\ding{55}}             & \ding{72}               & \ding{55}               & \multicolumn{1}{c|}{\ding{55}}             & \ding{55}               & \ding{55}               & \ding{55}             \\
\multicolumn{2}{c||}{\textbf{Minimize Tiles}}                                     & \ding{55}               & \ding{51}               & \multicolumn{1}{c|}{\ding{55}}             & \ding{55}               & \ding{51}               & \multicolumn{1}{c|}{\ding{55}}             & \ding{55}               & \ding{55}               & \multicolumn{1}{c|}{\ding{55}}             & \ding{55}               & \ding{109}               & \ding{55}             \\ \hline
\multicolumn{1}{c|}{\multirow{2}{*}{\textbf{Balance}}} & \textbf{Min. Steps} & \ding{55}               & \ding{55}               & \multicolumn{1}{c|}{\ding{55}}             & \ding{55}               & \ding{55}               & \multicolumn{1}{c|}{\ding{55}}             & \ding{55}               & \ding{55}               & \multicolumn{1}{c|}{\ding{51}}             & \ding{55}               & \ding{55}               & \ding{55}             \\
\multicolumn{1}{c|}{}                                  & \textbf{Min. Tiles} & \ding{55}               & \ding{55}               & \multicolumn{1}{c|}{\ding{55}}             & \ding{55}               & \ding{55}               & \multicolumn{1}{c|}{\ding{51}}             & \ding{55}               & \ding{55}               & \multicolumn{1}{c|}{\ding{55}}             & \ding{55}               & \ding{55}               & \ding{55}             \\ \hline \hline
\end{tabular}
\label{tab:heuristic_table_ticks_crosses}
\end{table*}


\subsubsection{Minimizing Steps and Tiles:}

Brute force always yields the true results for both minimizing tiles and minimizing steps at the cost of extremely high time complexity, making it impractical for large-scale circuits. Among heuristic approaches, the Greedy algorithm emerges as the closest performer to brute force in minimizing steps, with only a 7.5\% increase in the average number of tiles. Meanwhile, for minimizing tiles, Dynamic Programming achieves identical results to brute force but with significantly lower time complexity. However, this benefit comes at the expense of high space complexity. For users constrained by memory limitations, the best alternative to DP for minimizing tiles is the Random algorithm, which results in an average increase of 14.69\% in tile count compared to brute force but operates with lower space overhead.

\subsubsection{Balancing Number of Steps and Tiles:}

Our analysis reveals that every circuit exhibits two distinct Pareto-optimal fronts, each defined by a Pareto step value and a Pareto tile value. While both balanced optimization approaches — DP and Greedy —yield Pareto-optimal configurations, their strengths differ: DP minimizes the total number of tiles, whereas Greedy minimizes the total number of steps.

\subsubsection{Summary:}

Table~\ref{tab:heuristic_table_ticks_crosses} presents a general-purpose heuristic to guide users in selecting the appropriate optimization strategy based on three possible objectives. 
For a given optimization objective, an algorithm and optimization strategy marked with \ding{51} indicates that it provides the best possible results. If marked with \ding{72}, it yields results very close to true results and is a viable alternative if the best algorithm is not feasible. A marking of \ding{109} signifies the second-best alternative, which can be used when higher-ranked options are unavailable. Finally, if marked with \ding{55}, the strategy should not be used for that objective. Although our schemes were primarily designed with surface codes in mind, they can, in principle, be extended to other toric-code-based patches, such as Majorana surface-code patches~\cite{litinski2018lattice} or color-code patches~\cite{landahl2014quantum, bombin2006topological, kesselring2018boundaries}.

\section{Conclusion} \label{sec:conclusion}

In this work, we present a systematic and optimized framework for designing scalable quantum architectures that integrate data block layouts with magic state distillation protocols, addressing circuit-specific requirements and system constraints. 
We consider three optimization strategies: minimizing tiles to reduce qubit usage, minimizing processing steps to accelerate computation, and a balanced approach that achieves a trade-off between these metrics. We implement and analyze multiple optimization algorithms (\textit{random}, \textit{brute force}, \textit{dynamic programming}, and \textit{greedy}) to evaluate their performance across these strategies. To ensure comprehensive coverage, we examine a wide range of circuits. 
Our results reveal that minimizing steps shows a dependency on the number of columns while minimizing tiles correlates with the number of qubits. Comparative analysis shows that brute force delivers the best results for both steps and tiles but is computationally expensive. The greedy algorithm offers a close approximation for minimizing steps, deviating from brute force by only $7\%$, while dynamic programming matches brute force for minimizing tiles, offering an efficient alternative. Additionally, we explore Pareto-optimal trade-offs to balance steps and tiles. Finally, we propose a generalized heuristic to guide algorithm selection: the greedy algorithm is an effective alternative to brute force for minimizing steps, while dynamic programming—and occasionally the random algorithm—performs well for minimizing tiles. For balanced optimization, the greedy algorithm suits step-focused goals, and dynamic programming excels in tile-focused objectives. 

\section*{Acknowledgment}

The work is supported in parts by the National Science Foundation (NSF) (CNS-1722557, CCF-1718474) and gifts from Intel.

\bibliographystyle{unsrt}
\bibliography{refs}

\end{document}